\documentstyle[twocolumn,amssymb,aps,psfig]{revtex}
\tighten
\begin{document}
\draft

\title{\bf Minimal model to describe the magnetism of $CuGeO_{3}$
}

\author{G.~Bouzerar$^1$, \"O.~Legeza$^2$ and T.~Ziman$^3$}
\address{ 
$^1$Institut f\"ur Theoretische Physik, Frei Universit\"at Berlin,\\  
Arnimallee 14, D--14195 Berlin, Germany \\
$^2$ Research Institute for Solid State Physics and Optics,\\
H-1525 Budapest, P.\ O.\ Box 49, Hungary \\
$^3$Institut Laue Langevin \\
BP 156, 38042 Grenoble France\\
 }
\address{~ 
\parbox{14cm}{\rm
\medskip
We show that to describe properly the low energy 
excitations of $CuGeO_{3}$ 
one must include the effects of a transverse antiferromagnetic coupling, 
which is estimated to be  $J_{\perp}=0.15 J$. Owing to this coupling
 the frustration in the chains is significantly lower than recent 1D estimates
based on purely one-dimensional arguments, we find $J_{2}=0.2 J$.
 Furthermore we have found a strong modulation of the nearest
neighbour coupling due to the static distortion $\delta = 0.065$ ,
 which is 5 times higher than that previously deduced from a 1D chain approach.
Our set of parameters gives, i) a value of the distorsion which agrees well with some recent estimate for a lower bound, we are able to perfectly reproduce ii)the dispersions, iii) the experimental susceptibility at both high and low temperatures. By performing  DMRG  calculations for 2 coupled chains
we have analysed the effect of the transverse coupling on the ratio of
singlet to triplet gaps.
The ratio is very sensitive to the parameters and the universality
reported in the strict one dimensional case is lost.
As an additional point, we  provide a simple picture to explain the  interesting new feature observed in recent inelastic neutron scattering  experiments:
the existence of a second branch of excitations.
\\ \vskip0.05cm \medskip PACS numbers: 71.27.+a, 75.40.Mg, 75.90.+w
}}
\maketitle

\narrowtext

\section{Introduction}

The discovery of the non-organic compound $CuGeO_3$ has attracted considerable 
attention as a laboratory for low dimensional many-body quantum mechanics.
This compound is believed to exhibit a spin Peierls transition at $T=T_{Sp}$.
Below this temperature the ground-state is dimerized, and simultaneously a gap opens in the excitation spectrum.
Most of the attempts to extract the appropriate magnetic couplings
have been 
in the framework of purely one dimensional system.
This restriction to a one -dimensional picture was mainly justified by the features
of Inelastic Neutron Scattering, 
the dispersion is the largest in the chain direction (c-direction) 
\cite{NishiRegnault1}. 
The model which has been widely studied that with dimerization
and frustration;

\begin{equation}
H_{1D}=J_{c}\sum_{i}\left([1+\delta(-1)^i]{\bf S}_i\cdot{\bf S}_{i+1} + 
\alpha{\bf S}_i \cdot {\bf S}_{i+2}\right)
\label{hamilt}
\end{equation}

$\delta$ measure the distorsion of the lattice at zero temperature and 
$\alpha$ is the measure of frustration, the ratio of second-nearest to nearest
neighbour antiferromagnetic exchange. 
Attempts to fit the susceptibility data and the singlet-triplet gap have provided two different set of parameters $(J_c,\alpha,\delta)$:

a) $(J_c=150K,\alpha=0.24,\delta=0.030)$ in ref. \cite{Castilla} 

b) $(J_c=160K,\alpha=0.36,\delta=0.016)$ in ref. \cite{Riera}
The discrepancy in these two sets of parameters came from
differing emphasis in the fitting procedures: In reference \cite{Castilla}
the value $\alpha$ was first constrained to be at most 0.24 by the observation
that the triplet gap scaled with temperature as a power of the lattice
distortion and should thus be less than the (one-dimensional) values giving
a spontaneous dimerization. On the other hand it was made as large as possible
to give a reasonable fit to the susceptibility. The distorsion $\delta$ was 
fit from the dispersion at low temperatures. In reference \cite{Riera}
the value of $\alpha$ was simply chosen to make the best fit possible to the
susceptibility and $\delta$ was taken to fit the triple gap at zero
wave vector.

Recently, Fabricius et al. \cite{Fabricius} carefully  reexamined 
the fit of the  susceptibility to  a strictly one-dimensional model. They 
deduced  a value of $\alpha=0.354$,
which is almost identical to the  estimate of Riera and Dobry. Their fit  reproduced perfectly the susceptibility from $40 K$ up to $1000 K$.
For this reason the value $\alpha=0.36$ has been prefered to $\alpha=0.24$
and has been considered as {\it the} value from susceptibility measurements.

In fact  the issue is not settled, as apart from the observed relation of the
gap to the lattice distortion, Raman scattering
experiments lead to a value of $\alpha$ which is very much smaller, as we 
shall see: a
value of $\alpha$ close to 0.2 was suggested. The argument for this is as follows: 
It is known that the a system diescribed by (1) can exhibit a 
singlet bound-state excitation below the continuum \cite{Shastry-Uhrig}.
A recent detailed analysis have shown that it appears at $q=0$ momentum
 when the frustration is strictly non zero, for any given distorsion of the lattice\cite{Bouzerar} . 
The most interesting feature is that the ratio 
$R=\Delta_s/\Delta_t$ of the gap to singlet excitations measured in
Raman scattering and the gap to triplet excitations as visible
in inelastic neutron scattering was shown to be a universal function of the frustration parameter, i.e is independant of 
the dimerization parameter. This then, provides a direct tool to measure the 
amplitude of the frustration.
Singlet excitations are directly observable in  Raman scattering experiments.
Such a bound state was first observed by Kuroe et al. \cite{Kuroe}.
The excitation energy is $\Delta_s \approx 1.78 \Delta_t$ \cite{Loosdrecht}, it could be directly concluded from fig.1 in ref. \cite{Bouzerar} that this leads to $\alpha \approx 0.20$.
Note that this  value is very close to that proposed by Castilla et al\cite{Castilla}.
In contrast  a value of $\alpha \approx 0.35$ would give 
$R=1.50$, inconsistent with the experimental value.

In conclusion, a description purely in terms of uncoupled chains lead to a conflict:

i) Susceptibility data $\rightarrow $ $\alpha =0.36$.

ii) Raman scattering data $\rightarrow $ $\alpha \approx 0.2$.

The two values proposed suggest very different picture of the 
spin-Peierls transition: for a single chain there is a
critical $\alpha_{c}= 0.24$ below which dimerization
requires a coupling to phonons, and above which even without
coupling to the lattice the magnetic
chain spontaneously develops dimer order. While this may seem an
academic point, given that a lattice dimerization {\it is} observed,
but it is not if we wish to give a quantitative
account of the physics: the estimate of that exchange dimerization
depends on the frustration. 
It should be also stressed that Braden et al. have recently given an estimate of the lower bound value of the distorsion parameter wich has clearly excluded the value $\delta =0.016$ of scenario b) \cite{Braden}. Indeed they have found that $\delta > \delta_{min}=0.03$.
To solve this issue some have argued that an adiabatic 
approach was not appropriate and that the phonons degrees of freedom should be treated exactly ('dynamical phonon picture') \cite{Wellein}.
All recent attempts to include dynamical phonons were made in the framework of strictly 1D system.
Unfortunately, because of the large size of the Hilbert space, the  exact calculations are restricted to very small cluster\cite{Augier-Wellein}.

In the following we will show that if we take  into account properly 
the transverse coupling between chains, these fundamental issues are solved.
Within our approach, described in the next section, we will be able to reproduce perfectly many experimental features and provide a more realistic value value of the distorsion parameter $\delta$.

\section{Mimimal 2D model}

As was already mentioned in the first experimental papers,
$CuGeO_3$ is not a strictly one-dimensional compound: a na\"\i ve spin wave approach provide an estimate of the antiferromagnetic coupling in the perpendicular direction to the chain (b-direction), $J_{b} \approx 0.1 J_{c}$, which is not so small \cite{NishiRegnault1,NR2}.
Unfortunately, the problem with a two-dimensional system is
that much less is known both analytically and numerically.
Numerical methods such as  
Exact Diagonalization and Density Matrix Renormalization Group can resolve
only small clusters far from the thermodynamic limit.

Most  attempts  to study $CuGeO_3$ as a quasi 2D compound were essentially  within the framework of mean field theory \cite{2DMF}. An alternative, and powerful, approach for weakly coupled low-dimensional systems is to maintain a precise treatment of the one dimension and treat only the interchain coupling perturbatively. We shall 
follow this method and supplement it with exact treatment of two 
chains for certain parameters as a check on the numerical parameters found.
The low temperature phase of $CuGeO_3$ actually has a checker-board structure\cite{Hirota}: the dimerization alternates from chain to chain.
The minimal magnetic Hamiltonian is then:

\begin{equation}
H_{2D}=H_{1D}+J_{\perp}\sum_{i,r} {\bf S}_{i,r}\cdot{\bf S}_{i,r+1}
\label{hamilt2d}
\end{equation}

 with,

\begin{equation}
H_{1D}=J_{c}\sum_{i,r}\left([1+\delta(-1)^{i+r}]{\bf S}_{i,r}\cdot{\bf S}_{i+1,r} +
\alpha{\bf S}_{i,r} \cdot {\bf S}_{i+2,r}\right)
\end{equation}

The variable $r$  counts the different chains. The factor $(-1)^{r}$ explicitly takes into account the fact
that the structure of the real system is checker-board.

Starting from the limit of strong dimerization 
(in which triplet excitations  are on nearest neighbour sites), we can treat the coupling between the chains perturbatively.
It is easy to find, at the lowest order, that the dispersion is, 

\begin{equation}
\omega(k_{\parallel},k_{\perp})=\omega_{1D}(k_{\parallel})-J_{\perp}cos(k_{\parallel})cos(k_{\perp})
\label{dispers}
\end{equation}

where $k_{\parallel}$ is the wave-vector along the strongly coupled direction and $k_{\perp}$ is transverse.
Eq. \ref{dispers} implies the simple relation:
\begin{equation}
\Delta=\Delta_{1D}-J_{\perp}
\label{gap}
\end{equation}
$\Delta$ is the real gap (including $J_{\perp}$), and $\Delta_{1D}$ denotes the gap in absence of coupling between the chains.

Note that if the structure were not checker-board, but simply repeated in the b direction the dispersion would be different:
$\omega(k_{\parallel},k_{\perp})=\omega_{1D}(k_{\parallel})-
J_{\perp}cos(k_{\perp})$.

\par
While equation (4) is strictly true for very strong dimerization, ie $J_{\perp} <<  \omega (k_{\parallel})$ it is a very good approximation even when this is not satisfied, as we have verified by diagonalisation of two chains using 
Density Matrix Renormalization Group (DMRG).
We then exploit this equation and use the exact one-dimensional dispersion from exact diagonalisation $\omega_{1D}(k_{\parallel})$.
From exact diagonalisation, in the parameter region relevant for our discussion,  a good fit for the dispersion is \cite{Bouzerar.unpublished},

\begin{equation}
\omega_{1D}(k)=J_{c} \sqrt{a-bcos(2k)}
\label{dispers1d}
\end{equation}

where $a=\frac{1}{2}+\frac{\Delta_{1D}}{J_{c}}+(\frac{\Delta_{1D}}{J_{c}})^{2}$ and $b=\frac{1}{2}+\frac{\Delta_{1D}}{J_{c}}$.

\section{Fixing the parameters.}

In order to fix the parameters of the model, let us first start with the transverse coupling $J_{\perp}$.
For this purpose, we will use the experimental data for the dispersion as measured by inelastic neutron scattering \cite{NishiRegnault1}.
From equation (\ref{dispers}) the dispersion in the b-direction gives
 the transverse coupling from the difference beween the spin wave energy 
at $q_b$ the band edge and the centre, one has directly 

i) $ J_{\perp}= (5.6-2.0)/2=1.8 meV$

From equation (\ref{dispers})  the width of the 1D dispersion,
$\omega(\pi/2,k_{\perp})-\omega(0,k_{\perp})=\omega_{1D}(\pi/2)-\omega_{1D}(0) = \Delta \omega$. From eq. (\ref{dispers1d}) the width of the 1D dispersion is $\Delta \omega \approx J_{c}$. 
Thus using the given experimental data, we find,

ii)$ J_{c} \approx 12.2 meV= 146 K$

Hence, from i) and ii) we obtain $J_{\perp} \approx 0.15 J_{c} $, this value of the transverse coupling confirms the poor 1D character of $CuGeO_3$.
Note that the ratio of couplings is somewhat larger than
estimated by the experimentalists who used an expression appropriate
for a gap induced by anisotropy. As the measured anisotropy
in spin is small\cite{Brill} this is not appropriate.

Now our task is to fix the value of the two remaining parameters, $\alpha$ and
$\delta$.
As said before, this was  easy in the 1D picture, since the ratio singlet-triplet gap was shown to be a universal function of the frustration parameter only. 
Unfortunately, this is not the case anymore in the 2D 
model ($J_{\perp} \ne 0$). This will be illustrated later on.
An alternative  is to make use the high-temperature susceptibility data.
As a reference for the high-T data, we have recalculated for the one-dimensional system the susceptibility with $\alpha=0.35$ and $J_{c}=156 K$ and $g=2.25$,
since with this set of parameters, the experimental data have been perfectly reproduced from $40 K$ up to $1000 K$ \cite{Fabricius}.
It is easy to get the first terms in the high T expansion for the susceptibility, in this limit one gets straigthforwardly,

\begin{equation}
\chi (T)= \frac{N_{A}}{k_{B}T}(g \mu_{B})^{2} (a_{0}-\frac{a_{1}}{T})
\end{equation}
where $a_{0}=\frac{1}{4} $ 
and $a_{1}=\frac{J_{c}}{8}(1+\alpha+J_{\perp}/J_{c})$

For the purely 1D case, i.e $J_{\perp}=0$, one requires a value of $\alpha_{1D} = 0.35$. Thus the condition to reproduce the high T part of the susceptibility is,
\begin{equation}
\alpha_{reel} + {J_{\perp}\over J_c} = \alpha_{1D} \approx  0.35
\end{equation}

Since $J_{\perp}=0.15 J_{c}$, the previous equation fixes unambigously the amplitude of the frustration: 

iii) $\alpha_{reel}=0.20$

From this relation, it can be concluded that a purely 1D approach overestimates the real value of the frustration parameter. In other words, to reproduce the High T behavior of the susceptibility in the one-dimensional limit, one requires a larger value of the frustration parameter.

As a check, we have performed the exact calculation of the susceptibility for a two chains system. This was done for a $2 \times 8$ system, using periodic boundary conditions in both directions, this means for the two chain problem changing $J_{\perp}$ to 2 $J_{\perp}$.
For this size we can fully diagonalize the Hamiltonian in each subspace
and calculate the thermodynamic functions.
Since we are interested in the high temperature phase we have set $\delta=0$.
Note that, at high temperature the number of chains does not limit the accuracy of the calculation.
In fig.\ref{fig3} we see that the agreement between the experimental data and the high-T expansion is very good for sufficiently large temperatures.
 We also see that the agreement with the experimental data is very good down to 100 K.
One should not pay too much attention to the fact that the pure one-dimensional approach was very good down to $45 K$.
In our case, the finite size effect are definitely stronger:
the chains are shorter and at low temperature the number of chains is important
in the present case we have considered only two chains. Nevertheless the agreement goes beyond the strict applicablity of the leading term.

The last step consists in fixing the single remaining parameter, i.e the static distorsion amplitude.
For this purpose, we had to perform a calculation of the gap in the one-dimensional case, with the frustration parameter set to $\alpha=0.2$. 
Using eq. \ref{gap}, one immediately deduces $\Delta_{1D}=0.322 J_{c}$.
Following the method of ref. \cite{Bouzerar} to extrapolate the data in the thermodynamic limit, we have obtained straightforwardly, 

iv) $\delta \approx 0.065$. 

We remark that this value is significantly larger that previously reported: it is almost 5 times larger than the value obtained in the purely one-dimensional approach \cite{Riera} and closer to that estimated from the structure\cite{Braden}.
At this point all the parameters of the model have been fixed from the experimental data.
In order to test the validity of our set of parameters, we must confront it
to different tests. This will be done in the following section.

\section{Parameter tests.}
 
The realistic value of $\delta$ we have found is already a first check of the 
validity of our set of parameters. We now proceed to  a second test; let us check that we can accurately reproduce the dispersion data in both directions.
Using eq. (\ref{dispers}) and (\ref{dispers1d}) we have performed the direct 
calculation of the dispersion in the chain and transverse directions.
The plot of fig.\ref{fig1} is rather convincing, the agreement between experimental and theoretical data is excellent.

Now, as a third check of the validity of our approach, let us perform the calculation of the low temperature susceptibility ($T< T_{SP}$) and compare it to available experimental data.
For this purpose we assume that $\delta(T)=\delta$, which is reasonable
except very close to the transition point \cite{NishiRegnault1}.
Following the method of ref. \cite{Troyer} the susceptibility is accurately given by:

\begin{equation}
\chi(T)= \frac{N_{A}}{k_{B}T}(g \mu_{B})^{2} \frac{z(T)}{1+3z(T)}
\end{equation}
 whith,
\begin{equation}
z(T)=\frac{1}{(\pi)^2} \int dk_{\parallel}dk_{\perp} exp(-\frac{\omega(k)}{T})
\end{equation}

The expression for the dispersion is given by equation (\ref{dispers}).

We have plotted in fig. \ref{fig4} both the calculated susceptibility at low T and the experimental data \cite{Hase}. We observe that the agreement is very good up to $T \approx 0.85 T_{SP}$. 
Note that between $0.85 T_{SP}$ and $ T_{SP}$, one naturally expects a deviation from the experimental data due to the sudden drop of $\delta(T)$ when approaching the transition temperature. As 
a remark one has to keep in mind that we did not use any fitting parameter.

At this point, tt will be also an interesting point to estimate with our set of parameters the energy of the singlet bound state. As a preliminary step, 
let us first analyse the effect of the transverse coupling on the singlet-triplet ratio.
Indeed, as said in the introduction part, the one dimensionnal calculation have shown that this ratio does only depend on the frustration only, leading in this case to 
$\alpha \approx 0.2$, to get the experimental value $R_{exp} \approx 1.8$.
As we are now considering a weak, but non-zero transverse coupling we must know
whether this ratio will change substantially with interchain exchange.
To test this we have carried out Density Matrix Renormalization Group calculations on the system of two coupled chains.

\section {Effect of $J_{\perp}$ on the singlet-triplet ratio R.}

We have performed numerical calculations by applying the DMRG method 
\cite{white} on the model defined by the Hamiltonian in Eq.\ (\ref{hamilt2d}).
Since this method is more accurate for systems with free ends, we will 
consider our two coupled chain model with open boundary conditions
in the long direction. 
An unfavorable consequence of the DMRG  is that the total momentum
is not a good quantum number. As our aim is to determine the singlet 
and triplet energy gaps at $k=0$ momentum we have used the spin reflection 
symmetry to rule out excited levels belonging to other $k$ values \cite{legeza}. 
The singlet gap was calculated from the energy difference of the two lowest
lying energy levels in the $S_T^z=0$ spin sector with odd parity under spin reflection
while the triplet gap was obtained from the lowest levels of the $S_T^z=0$ and $S_T^z=1$
subspaces. 
Another  unfavorable consequence of open boundary conditions, as in the nearest-neighbor valence-bound
configuration of the bilinear biquadratic model \cite{vbs}, is that   
free $S=1/2$ spins remain at the ends of one of the chains, 
giving rise to a fourfold degenerate ground state. 
This is exactly the case for any finite chain length for the extreme dimerization limit
($\delta=1$) 
and is true asymptotically for $\delta<1$. 
Since the extra degeneracies make 
the analysis of the spectrum more difficult we have removed the two outmost spins
on one chain and set the end-coupling to $J=1+\delta$ on the other chain.
It worth  mentioning that the total symmetry of all states becomes opposite under
the spin reflection symmetry as
it had been  for the original problem. 
In most of the
calculations we have used the more accurate version of DMRG, the
so-called {\em finite-lattice method} to determine the energies more precisely.
To further improve the efficiency of the  
calculations, we have also included the left-right reflection symmetry 
and all the investigated states were targeted independently.
Since for finite dimerization the system is gapped, it was adequate to keep $100-200$ 
block states to have the truncation error below $10^{-5}$ for chain up to $100$ sites.
The absolute error of our calculation was estimated by 
comparing the ratio of the singlet and triplet energies $(R)$ obtained for
the ladder model using the DMRG procedure at $\alpha_c=0.24$, $J_{\perp}=0$ 
to the exact value
$R=\sqrt{3}$ given by the sine-gordon model. For stronger dimerization the deviation 
was $10^{-3}$ while for weaker dimerization it was found to be $10^{-2}$.

The results  for the ratio of the singlet and triplet energy gaps $(R)$ as 
a function of $J_{\perp}$ in the strong dimerization regime $\delta=0.2$) and $0.2<\alpha <0.26$ values are plotted in fig. \ref{fig2}. 
It is seen from the figure that there exists 
a finite  region of $J_{\perp}$ where the value of $R$ is below 2, thus we have
confirmed the 
the bound state is stable even in the 2D case for not too large interchain couplings.
As a second result, we have found
that the triplet gap is a linear function of $J_{\perp}$ in agreement with eq.
\ref{dispers}. In general the coefficient of this linear function
should be different from unity and indeed this is apparent from the
two chain results. In order to estimate parameters, however, we take
the linear coefficient as unity as the two chain estimate
is too heavily biased by finite ( transverse ) size: doubling
the transverse coupling to take into account the periodic boundary condition
clearly oversetimates the correlations in the two transverse directions.
 The nonlinear behavior of $R$ follows from
the nonlinear dependence of the the singlet gap on $J_{\perp}$. Taking the parameters 
for the frustration and $R$
obtained from the 1D chain calculations, namely $\alpha=0.2, R=1.80$, one finds the same 
value of $R$ for a different parameter set: $\alpha\simeq 0.26$ $J_{\perp}\simeq 0.12$.

In order to check the effect of dimerization in the case of 2D system we have performed
the similar calculations, but for a weaker dimerization parameter $\delta=0.065$.
The result for $R$ as a function of $J_{\perp}$ is also shown on the figure.
It is apparent from the curves that the $R$ depends on $\delta$, 
thus the ratio is no longer a universal 
function of the frustration parameter for agiven finite interchain coupling. On the other hand, we have found the triplet gap is linear function of the interchain coupling and the slope does not depend on the strength of the static distortion.

It is clear that for weaker dimerization the ratio gets very sensitive
to the interchain coupling. For a fixed frustration, increasing interchain
coupling leads to the disappearance of the bound state into the continuum.
The fact that the ratio exceeds 2 means only that this state is no longer well defined. For instance, we find that when $\alpha=0.2$ the singlet bound-state disappear into the continuum for $J_{perp} > 0.1$.
While the value of the ratio of singlet to continuum can no longer be used
to extract directily a value of the frustration, the very existence of 
the state gives a lower bound on the frustration.
To conclude this section, one can say that it is extremely difficult from this two chain analyse to reproduce the experimental value of the singlet-triplet ratio: one requires very precise values of the parameters of the model.



\section{Existence of a second branch in INS}

Up to now we have assumed that the magnetic structure is
of equivalent but coupled chains at high temperature and
below the spin-Peierls transition
an identical dimerization along each chain in a checker-board structure,
ie the dimerization out of phase from chain to chain.
There are two modifications to this view: one which we take 
to be unconfirmed, the other which we take to be reliable.
\par
The first is that
X-ray analysis of the room temperature structure
have suggested a larger unit cell\cite{Hidaka,Yamada} even above the spin-Peierls transition,
in contrast to that found by neutrons\cite{Braden}. 
Re-examination by neutron diffraction\cite{Braden98}
confirmed the earlier neutron results that these distortions are absent in the crystals
studied by neutron diffraction, so 
while the differences in the experimental structures remain to
be fully elucidated, we shall assume that the
structure at high temperatures is as taken.
\par
The second modification we take as necessary to take into account.
Recent inelastic neutron results\cite{Lorenzo99} on the dynamics have shown
a new magnetic mode, that the experimentalists  call an optic mode. We shall now show that
the dispersion and amplitudes of this second mode are well explained
by a slight refinement of the Hamiltonian considered so far
in supposing two inequivalent chains at low temperatures.
As the numerical differences are small, this alters very little
the dispersion and the susceptibility at low temperatures.
We now give a simplified version based on the limit of strong dimerization.

It is enough for our purpose to consider a system which consists of two inequivalent chains A and B coupled antiferromagnetically through $J_{\perp}$. We first define the following set of parameters $(J_{A}=J_{c}+dJ,\delta_{A}=\delta+d\delta,\alpha_{A}=\alpha+d\alpha)$ and $(J_{B}=J_{c}-dJ,\delta_{B}=\delta-d\delta,\alpha_{B}=\alpha-d\alpha)$ respectively for the chain A and B where it is assumed that dJ, $d\delta$ and $d\alpha$ are small.
We also define $\omega_{A}(k_{\parallel})$ and $\omega_{B}(k_{\parallel})$ as the excitation energy for each chain, and $\Psi_{A}(k_{\parallel})$, $\Psi_{B}(k_{\parallel})$ the associated eigenvectors  .
In the strong dimerization limit, it is relatively easy to perform the calculation of the eignemodes. For a given $k_{\parallel}$, there are two possible modes $\omega_{+}$ and $\omega_{-}$ corresponding to $k_{\perp}=0$, or $\pi$.

\begin{equation}
\omega_{+} (k_{\parallel})=\omega(k_{\parallel})+\sqrt{\delta \omega (k_{\parallel})^{2}+[J_{\perp}cos(k_{\parallel})]^{2}},
\end{equation}

\begin{equation}
\omega_{-} (k_{\parallel})=\omega(k_{\parallel})-\sqrt{\delta \omega (k_{\parallel})^{2}+[J_{\perp}cos(k_{\parallel})]^{2}}
\end{equation}

Their corresponding eigenstates are,

\begin{equation}
\Psi_{+}(k_{\parallel})=\Psi_{A}(k_{\parallel})+(1-\frac{\delta \omega (k_{\parallel})}{J_{\perp} cos(k_{\parallel})})\Psi_{B}(k_{\parallel}),
\end{equation}

\begin{equation}
\Psi_{-}(k_{\parallel})=\Psi_{A}(k_{\parallel})-(1+\frac{\delta \omega (k_{\parallel})}{J_{\perp} cos(k_{\parallel})})\Psi_{B}(k_{\parallel}),
\end{equation}

where $\omega=(\omega_{A}+\omega_{B})/2$ and $\delta \omega =(\omega_{A}-\omega_{B})/2$.
For simplicity we have assumed that we are only working in the vicinity of $k_{\parallel}=0$ or $\pi$ and that 
$ \delta \omega \ll J_{\perp} cos(k_{\parallel})$, which is reasonable since the chains are only weakly inequivalent.
Now the crucial question is: Will we effectively observe both branches in INS experiment?

To answer this question we have to calculate the matrix elements, $F_{+}= \langle \Psi_{+} \mid S(\vec{k}) \mid \Psi_{GS} \rangle$ and $F_{-}=\langle \Psi_{-} \mid S(\vec{k}) \mid \Psi_{GS} \rangle $, where $\mid \Psi_{GS} \rangle$ is the Ground-state wave function and $S(\vec{k})=S^{Z}_{A}(k_{\parallel})+e^{ik{\perp}b}S^{Z}_{B}(k_{\parallel})$. 
When performing the calculation at the lowest order one gets,

\begin{equation}
F_{+}=f_{0}(1+e^{ik{\perp}b}+\frac{\delta \omega (k_{\parallel})}{J_{\perp} cos(k_{\parallel})})-df_{0}(1-e^{ik{\perp}b})
\end{equation}

\begin{equation}
F_{-}=f_{0}(1-e^{ik{\perp}b}-\frac{\delta \omega (k_{\parallel})}{J_{\perp} cos(k_{\parallel})})-df_{0}(1+e^{ik{\perp}b})
\end{equation}

where $f_{0}=\langle \Psi(k_{\parallel}) \mid S^{Z}(k_{\parallel}) \mid \Psi_{GS} \rangle$ for an isolated chain, considering the 'average' parameters 
$(J_{c},\delta,\alpha)$, $df_{0}$ corresponds to its variation.
If we now fix for instance $k{\perp}=0$ then the spectral weight in both branches is,

\begin{equation}
A_{+}=\| f_{0}\| ^{2}(1+\frac{\delta \omega (k_{\parallel})}{2J_{\perp} cos(k_{\parallel})})^{2}
\end{equation}

\begin{equation}
A_{-}=\| f_{0}\| ^{2}(\frac{\delta \omega (k_{\parallel})}{2J_{\perp} cos(k_{\parallel})})^{2}+\| df_{0}\| ^{2}+\frac{\delta \omega (k_{\parallel})}{2J_{\perp} cos(k_{\parallel})}df_{0}f_{0}^{*}
\end{equation}

It is clear from these expressions that one will effectively observe two excitation branches, one with a significant spectral weight (the main branch) and another one with a relatively small one (secondary), in agreement with the experimental observation. 
 Typically the order of magnitude of the ratio of the spectral weight is,

\begin{equation}
\frac{A_{-}}{A_{+}} \approx [\frac{\delta \omega (k_{\parallel})}{2J_{\perp} cos(k_{\parallel})}]^{2}
\end{equation}
This very simple calculation can be refined to make more quantitative comparison with the available experimental data.
Let us repeat again that the previous simplified expressions are only valid in the vicinity of $k_{\parallel}=0$ or $\pi$ and we are considering only two chains: ie $k_{\perp}$ is by definition $0$ or $\pi$.

\section{Conclusion}

In this paper we have proposed a minimal model to describe the magnetism
of $CuGeO_{3}$. It is shown unambigously that the compound is two-dimensional (the coupling in the third direction is very small).
We clearly identify the origin of the conflicts in the one-dimensional 
approach: i)small value of $\delta$ and ii)large value of $\alpha$ estimated from the susceptibilty data not consistent with the Raman experiments.
The susceptibility previously calculated in a pure one-dimensional picture has in fact strongly overestimated the amplitude of the frustration.
The smallness of $\delta$ previously quoted, is simply due to underestimating the 1D gap
(see eq. \ref{gap})
As a test of our set of parameters, we now provide a value of the distorsion which is more realistic, we are also able to reproduce perfectly the high temperature (uniform phase) and low temperature (dimerized phase) behavior
of the susceptibility without any additional fitting parameter. 
We could also 
perfectly reproduce the dispersion in both directions.
As mentionned above, it is difficult to reproduce the singlet-triplet ratio 
from a study of two coupled chains, because of the extreme sensitivity of the ratio to the parameters. However, we believe that it would be interesting to perform the calculation of the Raman intensity, this will provide a measurable effect of the transverse coupling.
We believe that this model with the parameters 
b) $(J_c=146K,\alpha=0.2,\delta=0.065,J_b/J_c=0.15)$ 
should be the starting point for further studies on 
$CuGeO_{3}$. For instance, the need to consider two inequivalent chains,
as explained from observation of the extra branch
in the inelastic scattering, can be included by taking
two slightly different values of the parameter $\delta$.
\par
As we have remarked, a value of $\alpha$ well below the critical value for a chain implies that understanding of the dimerization inevitably involves the coupling to the lattice: the observed
lattice distortion is not a secondary effect. The change of estimate implies that calculation
of dilute phase diagram is quantitatively very different. While we consider the model
with these parameters as the correct minimal model consistent with what is currently known.
\par
Copper Germanate is more correctly considered as a (spatially) anisotropic
spin system rather than as a quasi-one-dimensional spin chain.
The one-dimensional analyses that have been used until now therefore needed
to be quantitatively modified. The apparent conflict between parameter
sets previously obtained from different experiments reflects
the simplification.
While we can give a satisfactory account to the experimental
results to date, their remain aspects that require
precision:
\begin{itemize}
\item{} A  more precise value of the interchain coupling requires a quantitative calculation in the anisotropic two dimensional spin system. Our simple 
estimate based on the strong dimer limit $J_b/J_c=0.15$ may be modified
by this.This also implies that $\alpha$ and $\delta$ will be also affected.

\item{} The method proposed\cite{Bouzerar} to fix the 
frustration from the ratio of the singlet excitation seen in Raman to 
the triplet in neutron scattering is correct for a strictly one-dimensional
model but must be modified in the two dimensional case. 
We have verified the linearity of the triple gap predicted by the extreme dimer limit
by means of a numerically exact two chain result. The existence
and the position of a bound state in the Raman spectrum
certainly constrain the parameters, but in a way which defies
precise calculation at present.
That is why, it may be useful to extend calculations of the Raman scattering to include the form of the continuum.
\item{} As yet there is no detailed understanding of anisotropic
terms in the Hamiltonian. The bulk susceptibilities differ\cite{Hase}
but this has been interpreted in terms of anisotropy in the
tensor of g factors.
This may change with better understanding of the polarized neutron 
scattering cross-sections and/or more quantitative theories of the 
ESR\cite{Nojiri}.
\item{} The observation of the second inelastic branch in the
neutron scattering indicates that the magnetic unit cell is
doubled by a small inequivalence of adjacent chains in the low temperature
phase. This is also seen in recent ESR experiments\cite{Nojiri}.It would be interesting to explore what physical parameter is
inequivalent.
\end{itemize}

\section{Acknowledgements}
We are grateful to Jean-Paul Boucher, Emilio Lorenzo, Beatrice Grenier  and Louis-Pierre Regnault for
discussion of unpublished experimental data and for Olivier C\'epas and
Andrew Wills for 
discussions of the structures and anisotropies. G. Bouzerar thanks Ph. Nozi\`eres for hospitality at the Institut Laue Langevin. \"Ors Legeza was partially supported by the Hungarian Research Fund (OTKA) Grant No. 30173 and the French Ministry of Research and Technology Fund.

%
%

\begin{figure}
\caption[]{
Dispersion in both directions. The symbols are experimental data from
 taken from ref.\cite{NishiRegnault1}.
}
\label{fig1}
\end{figure}

\begin{figure}
\caption[]{
Data for the effect of $J_{\perp}$ on the Singlet-Triplet ratio R.
The distorsion $\delta$ is fixed to 0.2 (a) and to 0.065 (b).
We have considered several values of the frustration parameter $\alpha$.}

\label{fig2} 
\end{figure}

\begin{figure}
\caption[]{
  Susceptibility at High Temperature. Indirect Comparison between high-T expansion and experimental data.''Reference'' was recalculated in the 1D picture for a $L=16$ sites system using the parameters of ref. \cite{Fabricius}.
}
\label{fig3}
\end{figure}

\begin{figure}
\caption[]{
Susceptibility at Low Temperature. We compare the experimental data \cite{Hase} with the analytical calculations.
}
\label{fig4}
\end{figure} 


\begin{references}

\bibitem{NishiRegnault1} M. Nishi, O. Fujita, and J. Akimitsu, Phys. Rev. B 
{\bf 50}, 6508 (1994); L.P. Regnault, M. Ain, B. 
Hennion, G. Dhalenne, and A. Revcolevschi, Phys. Rev. B {\bf 53}, 5579 (1996);
\bibitem{Castilla} G. Castilla, S. Chakravarty, and V.J. Emery, Phys. Rev. 
Lett. {\bf 75}, 1823 (1995).
\bibitem{Riera} J. Riera and A. Dobry, Phys. Rev. B{\bf 51}, 16098 (1995).
\bibitem{Fabricius} K. Fabricius, A. Kl\"umper, U. L\"ow, B. B\"uchner, T.
Lorenz, G. Dhalenne, and A. Revcolevschi, Phys. Rev. B {\bf 57}, 1102 (1998).
\bibitem{Shastry-Uhrig}B.S Shastry and B. Sutherland, Phys. Rev. Letter {\bf 47}, 964 (1981), G.S. Uhrig and H.J. Schulz, Phys. Rev. B {\bf 54}, R9624 (1996).
\bibitem{Bouzerar} G. Bouzerar, A.P. Kampf and G.I. Japaridze, Phys. Rev. B {\bf 58}, 3117 (1998).
\bibitem{Kuroe} H. Kuroe et al. , Phys. Rev. B{\bf 50}, 16468 (1994).
\bibitem{Loosdrecht} P.H.M. van Loosdrecht, J.P. Boucher, G. Martinez, G. Dhalenne and A. Revcolevschi, Phys. Rev. Lett. {\bf 76}, 311 (1996).
\bibitem{Braden} M. Braden, G. Wilkendorf, J. Lorenzana, M. A\"\i n, G. McIntyre, M. Behruzi, G. Dhalenne, and A. Revcolevschi, Phys. Rev. B {\bf 57}, 1105 (1996).
\bibitem{Wellein}B. B\"uchner, H. Feskhe, A.P. Kampf, and G. Wellein cond-mat 9806022.
\bibitem{Augier-Wellein} D. Augier, D. Poilblanc, E. Sorensen, I. Affleck,
Phys. Rev. B {\bf 58}, 9110 (1998);
G. Wellein, H. Fehske and A.P. Kampf Phys. Rev. Lett. {\bf 81}, 3956 (1998).
\bibitem{NR2}
L.P. Regnault, M. Ain,B. Hennion, G. Dhalenne, and A.
Revcolevschi, Physica B {\bf 213\& 214}, 278 (1995); 
M.C. Martin, G. Shirane, Y. Fujii, M. Nishi, O. Fujita, J. Akimitsu, M. Hase, 
and K. Uchinokura, Phys. Rev. B {\bf 53}, R14713 (1996).
\bibitem{2DMF}W. Brenig, Phys. Rev. B{\bf 56}, 14441 (1997); G.S. Uhrig,
Phys. Rev. Lett. {\bf 79}, 163 (1997); J.Zang, S. Chakravarty and A.R. Bishop,
Phys. Rev. B{\bf 55}, R14705 (1997);R. Werner, C. Gros, Phys. Rev. B {\bf 57}, 2897 (1998)
\bibitem{Hirota}K. Hirota, D.E. Cox, J.E. Lorenzo, G. Shirane, J.M. Tranqueda, M. Hase, K. Uchinokura, H. Kojima, Y. Shibuya, I. Tanaka, Phys. Rev. Lett. {\bf 73}, 736 (1994).
\bibitem{Bouzerar.unpublished} G. Bouzerar unpublished.
\bibitem{Brill} T. M. Brill, J. P. Boucher, J. Voiron,
 G. Dhalenne,  A. Revcolevschi, and J. P. Renard Phys. Rev. Lett. {\bf 73}, 1545 (1994).
\bibitem{Troyer} M. Troyer, H. Tsunetsugu and D. Wurtz , Phys. Rev. B {\bf 50}, 13515 (1994).
\bibitem{Hase} M. Hase, I. Terasaki, K. Uchinokura, Phys. Rev. Lett. {\bf 70}
3651 (1993).
\bibitem{white}S.\ R.\ White,
         Phys.\ Rev.\ Lett.\ {\bf 69}, 2863 (1992);
         Phys.\ Rev.\ B {\bf 48}, 10345 (1993).
\bibitem{legeza}\"O.\ Legeza, J.\ S\'olyom
         Phys.\ Rev.\ B {\bf 56},  14449 (1997).
\bibitem{vbs}I.\ Affleck, T.\ Kennedy, E.\ H.\ Lieb, and H.\ Tasaki,
         Phys.\ Rev.\ Lett. {\bf 59}, 799 (1987);
         Commun.\ Math.\ Phys. {\bf 115}, 477 (1988).
\bibitem{Hidaka} M. Hidaka, M. Hatae, I. Yamada,M. Nishi, J. Akimitsu, J. Phys. Condens. Matter.{\bf 9}, 809 (1997).
\bibitem{Yamada} I. Yamada, M. Hidaka, M. Nishi, J. Akimitsu, J. Mag. and Mag. Materials{\bf 177-181}, 703 (1998).
\bibitem{Braden98} M. Braden, E. Ressouche, B, B\"uchner, R. Kessler, G. Heger
 G. Dhalenne, and A. Revcolevschi, Phys. Rev. B {\bf 57}, 11497 (1998).
\bibitem{Lorenzo99}  J. E. Lorenzo, L.P. Regnault, J. P. Boucher,
B.Hennion,
G. Dhalenne, and A. Revcolevschi, EuroPhys.  Lett {\bf 45}, 619 (1999).
\bibitem{Nojiri} H. Nojiri, H. Ohta, S. Okubo, O. Fujita, J. Akimitsu, M. Motokawa cond-mat/9906074.
\end{references}
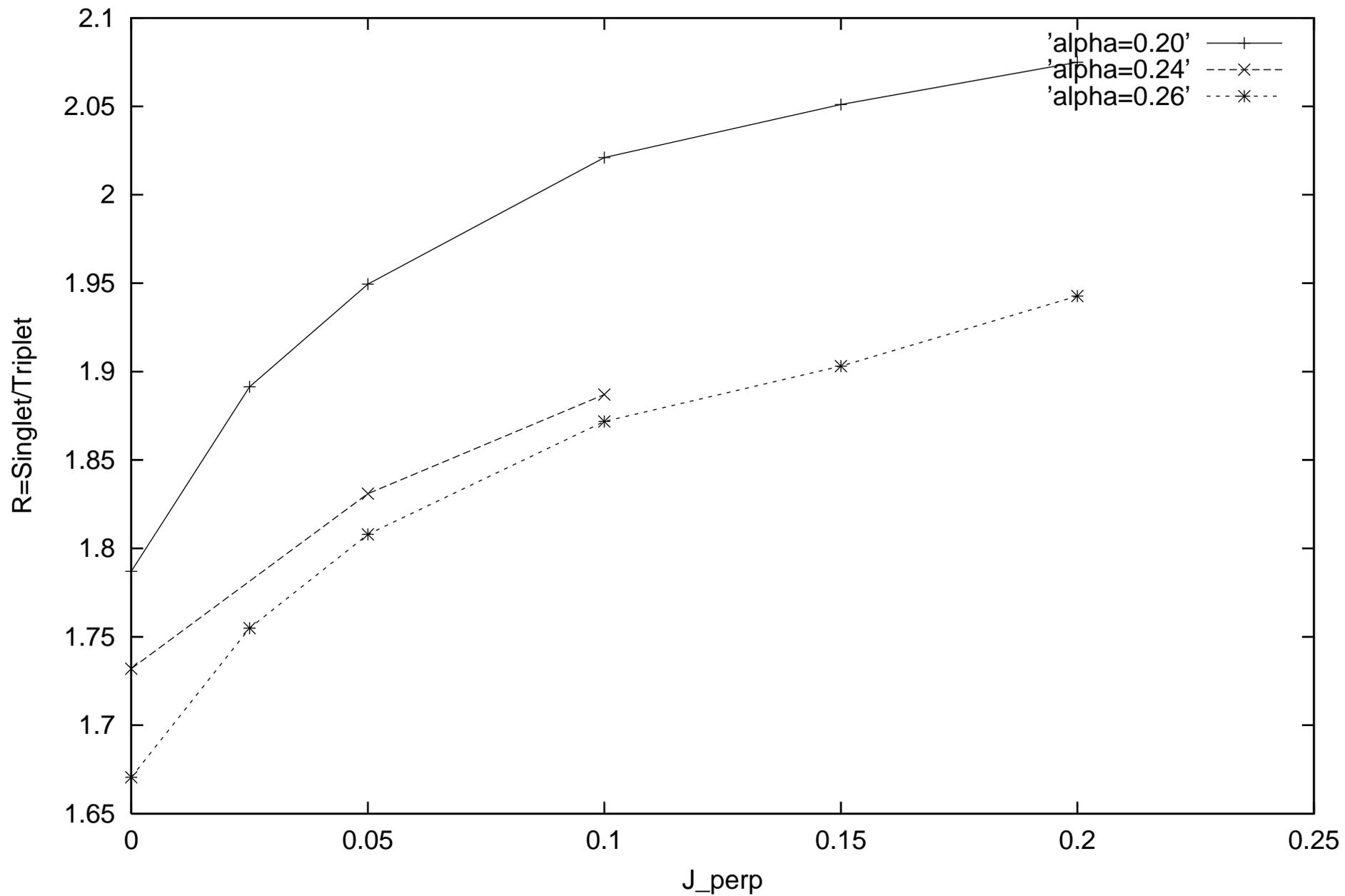
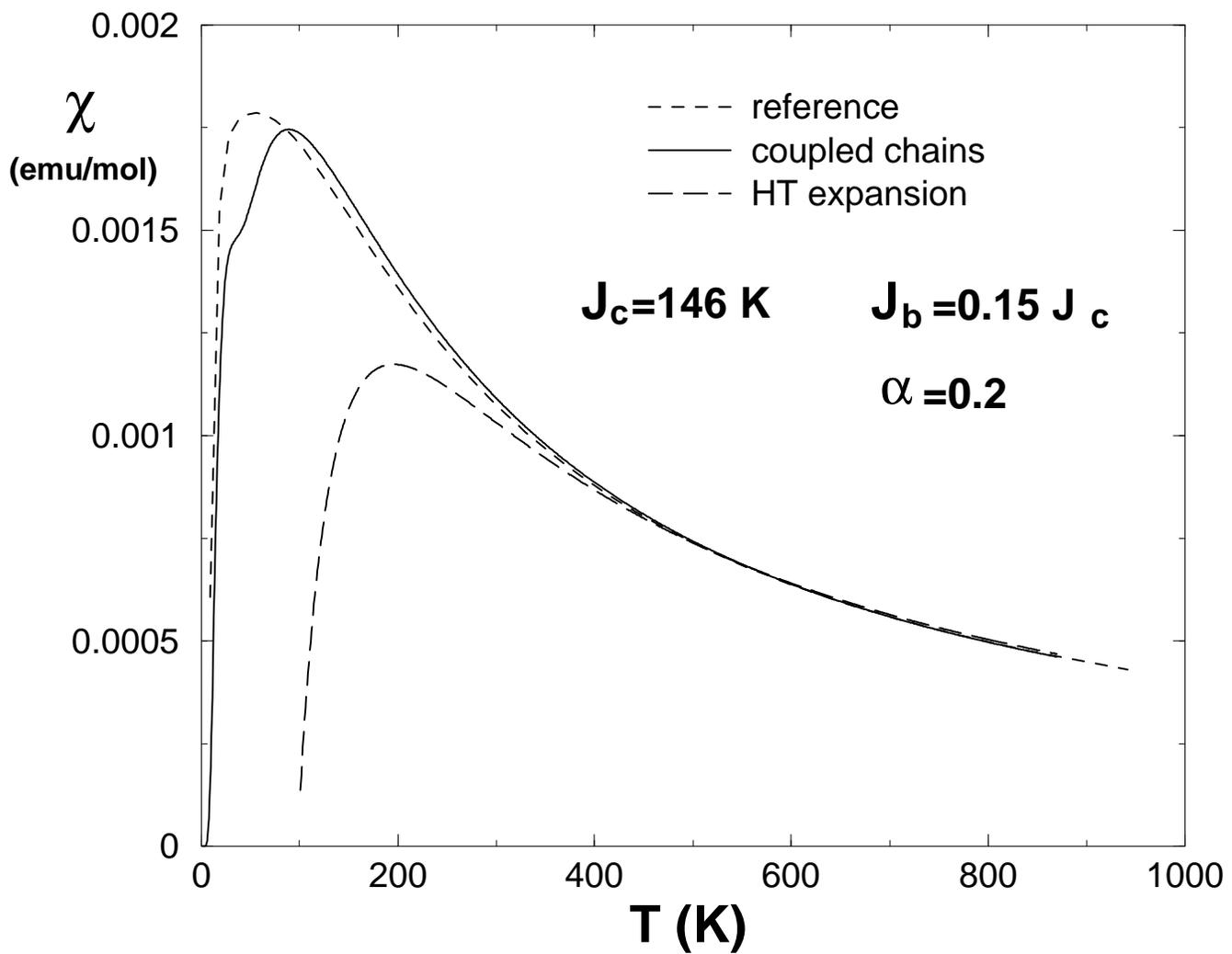
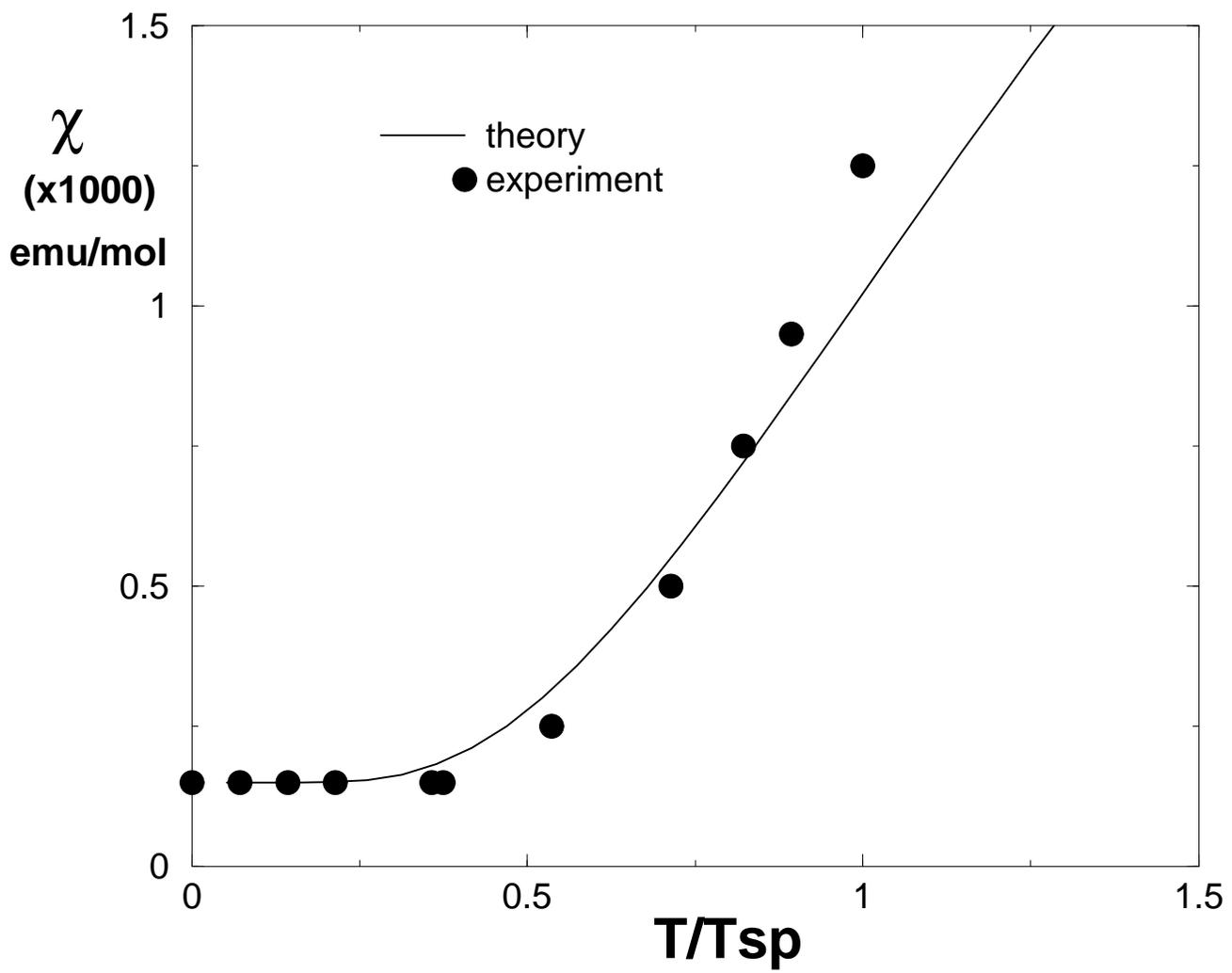
\end{document}